\newif\ifNotes\Notesfalse
\newif\ifAnon\Anonfalse
\documentclass[conference]{IEEEtran}

\newcommand{\swallow}[1]{}
\ifNotes
  \newcommand{\colorcomment}[2]{{\color{#1}[#2]}\xspace}
\else
  \newcommand{\colorcomment}[2]{\relax}
\fi

\newcommand{\urlfootnote}[1]{\footnote{\url{#1}}}

\usepackage{tikz}
\usepackage{amsmath}

\usepackage{cite}
\usepackage{doi}
\usepackage{amsmath,amssymb,amsfonts}
\usepackage{algorithmic}
\usepackage{graphicx}
\usepackage{textcomp}
\usepackage{array}
\usepackage{multirow,fixltx2e}
\usepackage{xspace}
\usepackage{hyperref}
\usepackage{color}
\usepackage{xcolor}
\let\labelindent\relax

\usepackage[inline]{enumitem}
\usepackage[nameinlink,capitalise,noabbrev]{cleveref}
\def\BibTeX{{\rm B\kern-.05em{\sc i\kern-.025em b}\kern-.08em
    T\kern-.1667em\lower.7ex\hbox{E}\kern-.125emX}}
\begin{document}

\definecolor{cyan}{rgb}{0.4,1,1}
\definecolor{orange}{rgb}{1,0.7,0}
\definecolor{dkgreen}{rgb}{0,0.6,0}
\definecolor{gray}{rgb}{0.5,0.5,0.5}
\definecolor{purple}{rgb}{0.58,0,0.82}
\newcommand{\taggedcolorcomment}[3]{\colorcomment{#1}{\textbf{#2}: #3}}
\newcommand{\todo}[1]{\colorcomment{red}{TODO: #1}}
\newcommand{\refs}{\colorcomment{red}{REFS}}
\newcommand{\AV}[1]{\taggedcolorcomment{brown}{Antoine}{#1}}
\newcommand{\pierre}[1]{\taggedcolorcomment{orange}{Pierre}{#1}}
\newcommand{\naif}[1]{\taggedcolorcomment{magenta}{Naif}{#1}}
\newcommand{\WR}[1]{\taggedcolorcomment{purple}{Walter}{#1}}
\ifNotes
  \newcommand{\hl}[1]{\textcolor{dkgreen}{#1}} 
  \newcommand{\hll}[1]{\textcolor{orange}{#1}} 
  \newcommand{\fix}[1]{\textcolor{red}{#1}} 
\else
  \newcommand{\hl}[1]{{#1}}
  \newcommand{\hll}[1]{{#1}}
  \newcommand{\fix}[1]{{#1}}
\fi
\newcommand{\parhead}[1]{\vspace{3pt plus 1pt minus 1pt}\par\noindent\textbf{#1}\hspace{.75em minus .5em}}
\newcommand{\etal}{\textit{et~al.\xspace}} 
\newcommand{\ie}{i.e.,\ } 
\newcommand{\eg}{e.g.,\ }
\ifAnon
\newcommand{\Datadome}{\colorbox{black!60}{\textcolor{white}{\textsc{Redacted}}}\xspace}
\else
\newcommand{\Datadome}{\textsc{Datadome}\xspace} 
\fi

\title{\Large \bf Free Proxies Unmasked: A Vulnerability and Longitudinal Analysis of Free Proxy Services}
\ifAnon
\author{Anonymous Authors*}
\else

\author{\IEEEauthorblockN{Naif Mehanna}
\IEEEauthorblockA{Univ. Lille, Inria, CNRS\\
naif.mehanna@univ-lille.fr}
\and
\IEEEauthorblockN{Walter Rudametkin}
\IEEEauthorblockA{Univ. Rennes, Inria, \\ CNRS, IRISA, IUF\\
walter.rudametkin@irisa.fr}
\and
\IEEEauthorblockN{Pierre Laperdrix}
\IEEEauthorblockA{CNRS, Univ. Lille, Inria\\
pierre.laperdrix@univ-lille.fr}
\and
\IEEEauthorblockN{Antoine Vastel}
\IEEEauthorblockA{Datadome\\
antoine.vastel@datadome.co}
}

\fi

\IEEEoverridecommandlockouts
\makeatletter\def\@IEEEpubidpullup{6.5\baselineskip}\makeatother
\IEEEpubid{\parbox{\columnwidth}{
    {\fontsize{7.5}{7.5}\selectfont Workshop on Measurements, Attacks, and Defenses for the Web (MADWeb) 2024 \\
    1 March 2024, San Diego, CA, USA \\
    ISBN 979-8-9894372-2-1 \\
    https://dx.doi.org/10.14722/madweb.2024.23035 \\
    www.ndss-symposium.org}
}
\hspace{\columnsep}\makebox[\columnwidth]{}}

\maketitle
\pagestyle{plain}

\begin{abstract}
    Free-proxies have been widespread since the early days of the Web, helping users bypass geo-blocked content and conceal their IP addresses. Various proxy providers promise faster Internet or increased privacy while advertising their lists comprised of hundreds of readily available free proxies.
    However, while paid proxy services advertise the support of encrypted connections and high stability, free proxies often lack such guarantees, making them prone to malicious activities such as eavesdropping or modifying content.
    Furthermore, there's a market that encourages exploiting devices to install proxies.
    
    In this paper, we present a 30-month longitudinal study analyzing the stability, security, and potential manipulation of free web proxies that we collected from 11 providers. 
    Our collection resulted in over $640,600$ proxies, that we cumulatively tested daily. 
    We find that only 34.5\% of proxies were active at least once during our tests, showcasing the general instability of free proxies. 
    Geographically, a majority of proxies originate from the US and China. 
    Leveraging the Shodan search engine, we identified $4,452$ distinct vulnerabilities on the proxies' IP addresses, including $1,755$ vulnerabilities that allow unauthorized remote code execution and $2,036$ that enable privilege escalation on the host device.
    Through the software analysis on the proxies' IP addresses, we find that $42,206$ of them appear to run on MikroTik routers.
    Worryingly, we also discovered $16,923$ proxies that manipulate content, indicating potential malicious intent by proxy owners. 
    Ultimately, our research reveals that the use of free web proxies poses significant risks to users' privacy and security. The instability, vulnerabilities, and potential for malicious actions uncovered in our analysis lead us to strongly caution users against relying on free proxies.
\end{abstract}

\section{Introduction}

Users are seeking more privacy on the Internet. At the same time, web proxies are being advertised as a valid privacy-enabling tool. Multiple reasons lead users to route their traffic through a proxy, such as promises of faster internet along with increased anonymity and a more private browsing experience. Financial reasons might come into play when choosing between VPNs and proxies, leading users to route their traffic through free proxies. These reasons are particularly motivated by the recent breaches of privacy and increasing data hoarding by various corporations and authorities~\cite{roesner2012detecting, acar2014web}.
Theoretically, proxies offer users a good alternative to other privacy-enhancing tools, such as VPNs.
Proxies provide the ability to circumvent content censorship, access geo-blocked content and anonymize user traffic while remaining easy to use for the regular user. 
\hl{VPNs on the other hand, benefit from the presence of end-to-end encryption, which provides better security but can lead to slower connection speeds compared to proxies.}
Proxies come in various types and shapes, each providing different degrees of advantages, and in consequence, different degrees of privacy guarantees. For instance, paid proxies, typically offer access to encrypted residential or datacenter proxies with high bandwidth and are only accessible to a handful of users. Some services might offer fully private proxies, guaranteeing that the user is the only one using the relay.
On the other hand, free proxies are made publicly accessible on the Internet to any user. However, it is widely assessed that freely available proxies provide little to no privacy guarantees and are usually used as honeypots for malicious content manipulation or eavesdropping on the user's traffic. 
These claims have been covered in recent research, further proving that free proxies are often operated for malicious purposes, altering content and even forcing malware on the user~\cite{mani2018extensive,tsirantonakis2018large}.
Yet, it is unclear when connecting to a free proxy if its behavior is malicious by nature or if it was compromised by a vulnerability without the administrator being aware of it.

In this paper, we shed light on the security of free proxies by identifying the software and vulnerabilities on the proxies' IP addresses through the Shodan database.
In particular, we present an overview of the most common software and hardware identified in IP addresses linked to proxies and identify their vulnerabilities, as well as whether they are exploitable and might have led to the creation of a malicious proxy. 
More specifically, we make the following contributions:

\begin{enumerate}
    \item We perform a longitudinal analysis of over $640,600$ free proxies that were collected daily from $11$ proxy providers over the duration of $30$ months. We study their diversity, their stability, and assess their reliability. In particular, we show that the free proxy ecosystem is diverse, with proxies ranging in protocols, location, and autonomous systems. We also emphasize that free proxies are part of a dynamic ecosystem that constantly provides new entries. Finally, we show that free proxies are generally unstable, with only a few showcasing consistent responses to our probes.
	\item We analyze the security of free proxies using the Shodan search engine. More particularly, we aim to identify the software and hardware that powers the collected proxies and the vulnerabilities that could be exploited by malicious attackers on those systems. \hl{We show that the presence of critical vulnerablities might have led to the proxies' creation and that some proxies are powered by deprecated and vulnerable software, which can lead to traffic eavesdropping and content alteration.}
	We identify a total of $4,452$ distinct \textit{Common Vulnerabilities and Exposures (CVEs)}, with $578$ showcasing a CVSS score of 9 and over. When crossing the CVEs with the CAPEC database,\urlfootnote{https://capec.mitre.org/} we find that $1,755$ of these vulnerabilities allow attackers to execute unauthorized commands and $2,036$ of them can help gain privileges on the vulnerable device.
	Most notably, we find that most of the collected proxies can be inferred to run on networks that include a MikroTik router. We also find various occurrences of proxies running on the same port where a connected camera was identified.
\end{enumerate}

The remainder of this paper is structured as follows. Section \ref{sec:background} provides an
overview of the state of the art. Section \ref{sec:methodo} presents our experimental methodology. Section \ref{sec:study} presents the results of our experiments. Section \ref{sec:discussion} discusses our results and the limitations of this study. Finally, section \ref{sec:conclusion} concludes this paper.

\section{Background and related work}\label{sec:background}

Proxies, whether they are distributed for free or for a financial incentive, are widely available over the Internet. Specialized websites advertising free proxies, and ranging widely in quality and quantity, have emerged over the years. These websites are mostly organized in the same way: they provide a list that enumerates the available proxies and their information, such as the protocol, the IP, and the used port. Some websites provide additional data such as the source country of the proxy, its latency, and its last known status. The widespread availability of such websites makes it fairly easy to have access to a consequent amount of proxies without requiring substantial effort from the user.

\subsection{Proxies}

\parhead{Definition.}
A web proxy is one of the most well-known types of network relays.
It acts as an intermediary between a client and a server where its most basic function is to \textit{relay} the network packets, but it also can be used to filter requests, improve performance by caching resources, and even bypass censorship.
The popularity of proxies comes from the fact that they are easy to use.
By using a small script, a browser extension, or by entering a few numbers in the settings of an operating system, proxies are easy to install and activate.
Proxies are characterized by several features: their country of origin, their IP addresses and ports, their latency, their bandwidth, and the protocol they support.
Their quality often depends on their source and whether they were obtained through a paid or free service.

\parhead{Free proxies.}
Since they are publicly advertised, free proxies will often be under heavy utilization as many users will try to access them at the same time, and might be used for both benign and malicious traffic.
Specialized websites have emerged over the years, providing extensive lists of free proxies, and while the ecosystem is growing, only a small fraction of the announced proxies actually works, as reported by Perino~\etal~with their ProxyTorrent study~\cite{perino2018proxytorrent}.
On the 180K proxies, they tested over a 10-months period, less than 2\% were indeed proxying traffic and about 10\% of them exhibited malicious behaviors by injecting ads or intercepting TLS connections.
In their study, they also developed a web extension, named Ciao, through which users could select a secure free proxy suiting their constraints (location, protocol, anonymity level, etc.) based on a series of tests aimed to identify malicious behavior in free proxies. 
Using their extension, ProxyTorrent provides unique insights into free proxies' users and their preferences.

Mani~\etal{}~\cite{mani2018extensive} follow the same methodology in their proxy analysis and provide insights obtained through a 50-day study of over $107,000$ open proxies. They found that even though the open proxies may be geographically diverse, only a few countries account for the majority of working proxies. Another key finding in their study shows that in most cases, the manipulated content is not malicious. However, when a free proxy owner has malicious motives, they can go as far as introducing cryptocurrency mining scripts and forcing trojan executables on the users. Mani~\etal also enriches their study by analyzing content manipulation over a Tor network.

Scott~\etal{}~\cite{scott2015understanding} analyzed thousands of free proxies and found that most of them were short-lived with a median life of 7 days and their users spanned legitimate organizations to automatic and malicious traffic.
A study by Choi~\etal{}~\cite{choi2020understanding} presented the largest open proxies analysis to date, collecting over $7,000,000$ proxies, of which over $1,000,000$ are open proxies. 
They observed that $28.23$\% of open proxies were used for spam and $6.97$\% participated in launching malicious attacks.
Finally, Tsirantonakis~\etal~\cite{tsirantonakis2018large} perform a large-scale analysis of open proxies involving $65,000$ proxies over a two-month period. They find that over $38$\% of working proxies perform some kind of content modification. However, confirming the results by Mani~\etal~\cite{mani2018extensive}, they stress that only a small subset of these proxies do in fact alter the expected content in a malicious way.

\parhead{Paid and residential proxies.}
Paying for an online proxy helps secure a more reliable infrastructure, with high availability and better bandwidth compared to a free one.
While most paid proxies offer servers with IP addresses from data centers around the world, some services offer a premium resource called residential IP addresses.
These addresses belong to Internet Service Providers (ISP) and are often assigned to homes and small businesses. 
This difference is crucial as traffic from an internet customer at home appears more legitimate than traffic coming from a data center.
IP address reputation is often used by protection companies to assess if one connection should be let through or not and a residential IP address is less likely to be blocked than one coming from a data center.
Mi~\etal{}~\cite{mi2019resident} investigated 5 Residential IP providers and found that a lot of traffic coming from them involved ad clicking, promotion, and malicious activities.
Moreover, more than 9\% of traffic destinations were detected as malicious showing how these services are used by malicious entities.
In their study, Chiapponi~\etal~\cite{chiapponi2022badpass} showed that residential IP addresses were particularly interesting to bot owners, which introduces more challenges for bot detection services as they cannot rely on the IP address's reputation. 
Finally, Tosun~\etal~\cite{tosun2021resip} look into the recruitment process of residential proxies providers and introduces a detection mechanism for the identification of unwanted proxies on personal devices.

\parhead{Protocols.}
Web proxies \hl{mainly} rely on two different protocols, namely HTTP and SOCKS.
HTTP proxies are designed exclusively for HTTP connections while SOCKS proxies can handle any underlying protocol without any limitations. 
They both have their own strengths and weaknesses in terms of performance and reliability and both protocols are offered by free and paid web proxies.

\subsection{Shodan search engine}
\label{subsec:shodan_background}

Conventional search engines, such as Google, DuckDuckGo, or Bing, allow users to search for websites that match their search criteria. As a result, these search engines crawl the visible web for websites that they categorize based on their content and keywords.
However, websites represent only a subset of the services accessible on the Internet, and much of the remaining part of the Internet remains hidden from inexperienced users.

In 2009, John Matherly launched Shodan, with the aim of identifying and indexing devices facing the Internet. Through a user-friendly graphical interface, the Shodan search engine allows users to search for specific IP addresses and browse through the Internet-facing services it has identified.
Shodan builds its database through large-scale crawls of available IP ranges. Upon discovering an openly accessible IP address, Shodan interrogates and fingerprints endpoints associated with known services. This information is then indexed and openly provided through the Shodan webpage.\footnote{\url{https://shodan.io/}} If an IP address is present in their database, the Shodan search engine displays related information such as the IP address's location, the organization associated with the IP address, service banners describing identified software, and the potential vulnerabilities linked with those services.
Shodan also attaches an optional tag to some crawled IP addresses, indicating specific characteristics that were identified. Tags might signal that an IP address has been used for Command \& Control (C2) attacks through the \textit{c2} tag or that an IP address is part of the Tor network through the \textit{onion} tag.

Shodan also offers an API that users can programmatically request to obtain the information displayed on the website. While Shodan was the pioneer in providing such a service, alternative platforms have been developed, such as Censys,\footnote{\url{https://censys.com/}} introduced in 2015, which offers similar services. Another more recent alternative is \textit{Hunter.how}.\footnote{\url{https://hunter.how/}}

Shodan has been extensively used in the literature, mostly to perform vulnerability assessments. In 2014, Bodenheim~\etal~\cite{bodenheim2014evaluation} evaluated the indexing and querying capabilities of Shodan for industrial control device identification. They find that Shodan is able to identify and index their testbed within $19$ days of exposing them to the Internet. A comparative analysis between Censys and Shodan by Bennett~\etal~\cite{bennett2021empirical} further evaluates the responsiveness of such services by measuring their update frequency: they found that both Shodan and Censys perform less than 40 banner grabs per month, showcasing a low resource impact while also ensuring regular updates to their data.
In 2018, Bugeja~\etal~\cite{bugeja2018investigation} leveraged the Shodan search engine to identify vulnerabilities in Internet-facing smart cameras. Their results show that a significant share of smart cameras presents serious known vulnerabilities, which are listed in the \textit{Common Vulnerabilities and Exposure (CVE)} database.
Finally, Albataineh~\etal~\cite{albataineh2019iot} used a set of queries crafted for the Shodan database to unveil the vulnerabilities of various Internet-facing devices. They find that numerous devices remain vulnerable to simple attacks due to the use of default credentials.

\section{Methodology}
\label{sec:methodo}

To efficiently collect, probe, and monitor the bot activity related to open proxies, we have designed a methodology\footnote{The research artifact accompanying this paper can be found at \url{https://github.com/naifmeh/free_proxies_unmasked}} for collecting \textit{HTTP} and \textit{SOCKS} proxies from various web providers, testing them using one \textit{honeysite} under our control and finally collecting and parsing security-related information through the Shodan API. The obtained insights are then presented in Section~\ref{sec:study}.

\subsection{Collecting the proxies}
\label{subsec:proxycollect}

In order to build our proxy database, we first gather a list of different proxy aggregators websites that provide both \textit{HTTP} and \textit{SOCKS} proxies. We collect the different websites that come up on top of our search result for the terms \textit{"free proxies"} and \textit{"open proxies"} on the \textit{Google} search engine in \hl{April 2021}. \naif{verify} 
For each of the proposed websites, we verify whether the proxies are truly freely available and not behind a paywall or conditioned on having an account. We also verify that the website does not take an aggressive bot-detection approach, which would potentially prevent our scraper from accessing the list. Finally, we try to select websites that tend to frequently provide new proxies, as opposed to websites with a low renewal frequency.
After the previous considerations, we selected 9 initial proxy providers out of the different proxy lists. \\
Some aggregators stopped providing proxies during our tests. As a result, following the same methodology, we identified and added \textit{GeoNode}\urlfootnote{https://geonode.com/} and \textit{AdvancedName}\urlfootnote{https://advanced.name/} in March 2023. \hl{Since the existing aggregators continue to supply new proxies over the duration of the study, introducing new proxy providers does not impact any result presented in Section~\ref{sec:study}. The new providers enrich our database with more potentially unique proxies, and their impact can be compared to improved performances from existing aggregators.} \naif{added following R1's comment} Table~\ref{table:proxyproviders} presents the proxy providers, along with their proxy list's URL and activity period. 

\begin{table}[]
	\centering
    \caption{List of proxy providers and their proxy list's URL}
    \renewcommand{\arraystretch}{1.3}
    \begin{tabular}{|c|c|c|}
    \hline
    \textbf{Proxy provider} & \textbf{Proxy list's URL}       & \textbf{Activity period}   \\ \hline
    Foxtools                & https://foxtools.ru             & April 2021 --- October 2022  \\ \hline
    FreeProxyList           & https://www.free-proxy-list.net & April 2021 --- November 2023 \\ \hline
    FreeProxyLists          & https://www.freeproxylists.net  & April 2021 --- August 2023 \\ \hline
    ProxyScrape             & https://api.proxyscrape.com     & April 2021 --- November 2023 \\ \hline
    HideMyName              & https://hidemy.name             & May 2021 --- November 2023 \\ \hline
    OpenProxy               & https://openproxy.space         & April 2021 --- November 2023 \\ \hline
    ProxyRack               & https://nntime.com              & April 2021 --- October 2021\\ \hline
    My Proxy                & https://www.proxyrack.com       & April 2021 --- November 2023 \\ \hline
    NNTime                  & https://www.my-proxy.com        & April 2021 --- September 2023\\ \hline
    GeoNode                 & https://proxylist.geonode.com/  & March 2023 --- August 2023\\ \hline
    AdvancedName            & https://advanced.name/          & March 2023 --- November 2023\\ \hline
    \end{tabular}
    \label{table:proxyproviders}
\end{table}

We use a \textit{Redis}\urlfootnote{https://redis.io/} queue for scheduling our crawls, which are performed using \textit{Puppeteer}\urlfootnote{https://pptr.dev} on the \textit{Node.JS} language. For each website, we identify and use the proxy table's CSS selector to collect its content and save the result in our database. On each crawl, we collect the proxies' IP addresses, their ports, and their protocol (HTTP or SOCKS).

Furthermore, we use the \textit{Maxmind GeoIP}\urlfootnote{https://www.maxmind.com/} database to identify the proxies' origin country and \textit{Autonomous Systems (AS)}.
We schedule the \textit{Redis} queue to run exactly once a day. The whole collection process is made to be non-blocking, meaning that if a proxy provider is not available for the daily collection, it does not impact the collection on the remaining providers, nor prevent existing proxies from being tested.
As of the end of October 2023, our crawling process resulted in a total of $640,693$ unique proxies collected. \naif{verify}

\subsection{Testing the proxies}
\label{subsec:proxyprobe}

In order to establish a detailed profile of the open proxy ecosystem, we design a testing methodology that is both scalable and failure-safe. We emphasize the scalability of our architecture as the number of collected proxies grows daily. 
We chose to test the proxies once every day: therefore, each run has to take less than 24 hours to complete.
In order to address the scalability of the system, we build our testing system on top of \textit{Bee-Queue's}\urlfootnote{https://github.com/bee-queue/bee-queue} Redis-based queue system. We implement two queues for the testing process: 
\begin{enumerate}
    \item The first queue schedules the complete round of testing and is set to execute only one process at a time. We chose to use a queue for this to handle the case where the testing process potentially takes more than 24 hours. In this case, the next run is not canceled but simply queued.
    \item The second queue tests each individual proxy, in a process that we describe later in this section. We configure Bee-Queue to perform a maximum of 320 tests in parallel. This allows us to avoid overloading the network while maintaining a significant testing speed.
\end{enumerate}

Each test requests the HTTP endpoint of the \textit{honeysite} under our control. The tests are performed using a simple GET request with a custom user-agent mimicking an up-to-date real browser. As some proxies tend to be slow and resolve after a long time, we set a timeout of 60 seconds for each test. Our timeout value aligns with the value chosen in related works~\cite{perino2018proxytorrent, tsirantonakis2018large}, which use timeouts ranging from 45 seconds to 180 seconds. 

If the request reaches the \textit{honeysite} and the proxy responds before the timeout, we consider the proxy as active. In this case, we perform two more tests:
\begin{itemize}
    \item We first verify that the returned content is identical to the expected content of our \textit{honeysite}, \hl{which contains a simple chain of characters. To this end, we perform a simple difference check between the expected and returned content.}  If the returned content is different, the proxy is flagged as a content manipulator. The received content is then saved to disk. As verifying the nature of content modification of open proxies is out of the scope of this paper, we do not perform a more advanced analysis of the returned content.
    \item We perform a second request to our \textit{honeysite} but this time on the \textit{HTTPS} endpoint as an attempt to verify whether the proxy works using the secure protocol. This test is performed under the same conditions as the test to the non-secure endpoint.
\end{itemize}

We divide the tests by separating the proxies into two batches depending on their protocol: \textit{HTTP} proxies are tested first, followed by \textit{SOCKS} proxies.

Tsirantonakis~\etal~\cite{tsirantonakis2018large} implement a strategy to dispose of inactive proxies after some time. Their choice is motivated by the need to test the same proxies multiple times a day. For our work, we chose to not implement such a strategy for two reasons: first, we chose to test the proxies only once a day as our study spans over $30$ months. Second, we believe that inactive proxies might be reactivated after some time and that this behavior is specific to each individual proxy, meaning that if we chose to remove proxies from our testing queue after a fixed number of days, we will not be able to monitor them if they resume activity.

As an attempt to avoid blocking errors, the testing queue is configured to resume in case of a system shutdown, given that enough time remains for it to complete without significantly impacting the next run. Additionally, we set alerts in case the number of tested proxies is significantly lower than expected. 
In total, through the $949$ runs of our observation period, we discarded \hl{29} runs that were subject to crashes or inconsistencies. Most crashes were caused by factors out of our control, such as Internet shutdowns, resource limitations, or hard-drive failures. \hl{It is important to note that the discarded runs were uniformly distributed throughout the study, occurring between 243rd and 928th runs, with a maximum duration of interruption of 7 consecutive days.} 

\subsection{Collecting Shodan data}
\label{subsec:shodan_collection}

We use the Shodan API to collect information about each collected proxy. If multiple proxies are using the same IP address and different ports, we only request the Shodan API once for this IP address and use the information for all concerned proxies. \hl{All Shodan data is collected only once, at the end of the study (October 2023)}. 
The returned data consists of a JSON object that includes ownership information, location, and identified services metadata (through Shodan's banners\footnote{\url{https://blog.shodan.io/what-is-a-banner/}}). As mentioned in Section~\ref{subsec:shodan_background}, Shodan also provides vulnerability information for a given service if applicable, using the CVE database. Whenever possible, the collected services are identified through the \textit{Common Platform Enumeration (CPE)} naming scheme.

We use the OpenCVE project\footnote{\url{https://www.opencve.io/}} to further enrich the CVE information with the associated \textit{Common Weakness Enumeration (CWE)} database and detailed \textit{Common Vulnerability Scoring System (CVSS)} information.

\section{Longitudinal study}
\label{sec:study}

As mentioned in Section~\ref{sec:methodo}, we collected proxies from 11 public providers for a duration of 30 months, ranging from April 2021 to the end of October 2023. 
Each collection was followed by a series of tests for each proxy in the database that included verifying the proxy's HTTPS support and testing whether the proxy tampers with the content. 
Whenever available, we collected information related to the proxies' IP addresses from the Shodan database.
Over the 30 months of data collection, we collected a cumulative total of $640,693$ proxies, which we tested daily, with an average of $3.36\%$ verified working proxies per run. In total, $221,319$ proxies passed the verification test at least once, representing a total of $34.5\%$ being active at least once during the duration of the study. 
Among the proxies, we identified \hl{$531,998$} unique IP addresses. For each IP address, we requested the associated data on the Shodan database, resulting in $232,348$ entries.

In this section, we present the observations of our longitudinal study and insights obtained from the Shodan database.

\subsection{Characterization}
\label{subsec:activity}

\begin{figure}[]
    \centering
    \includegraphics[width=1.1\columnwidth]{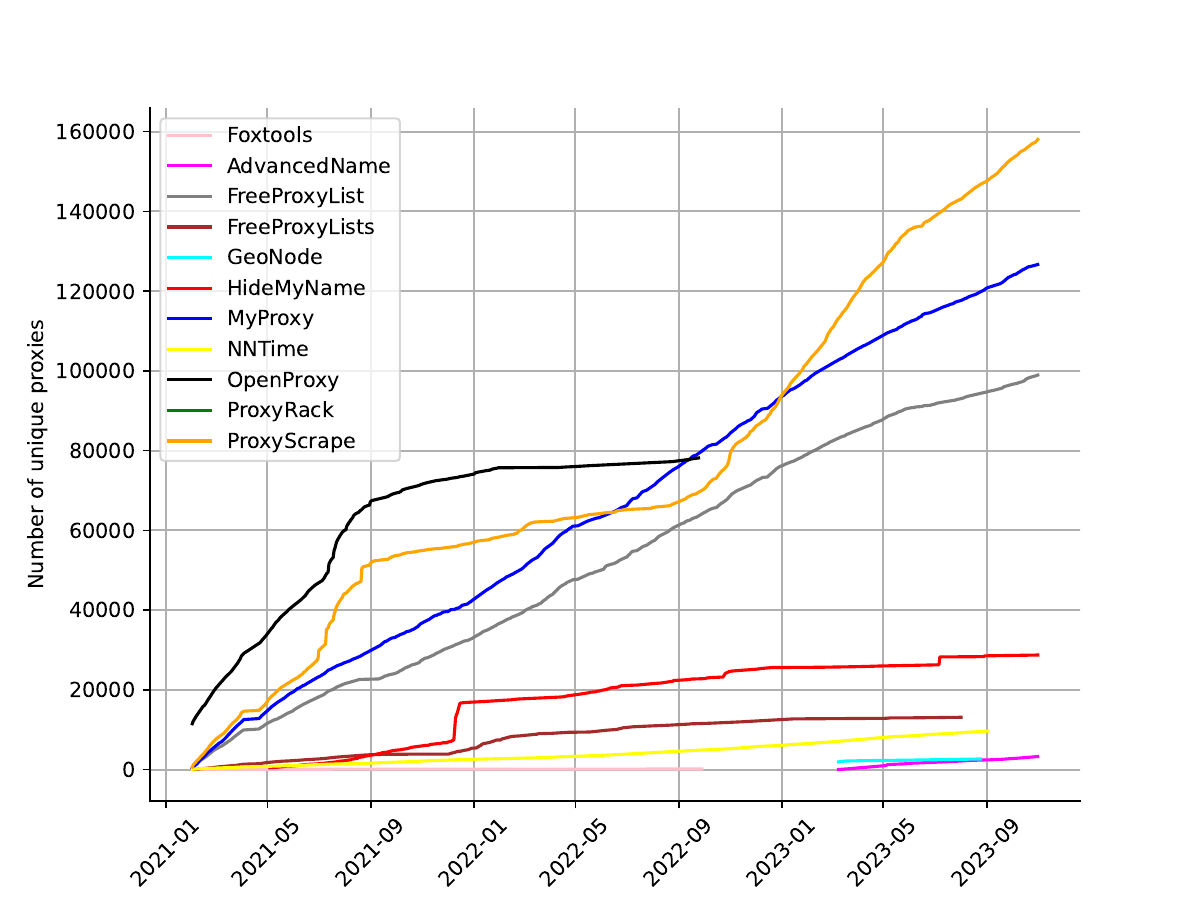}
    \caption{Cumulated sum of unique proxies per provider. A proxy is considered unique when it is only detected on one aggregator and has never been observed before.}
    \label{fig:evproviderunique}
\end{figure}

\begin{figure}[]
    \centering
    \includegraphics[width=1\columnwidth]{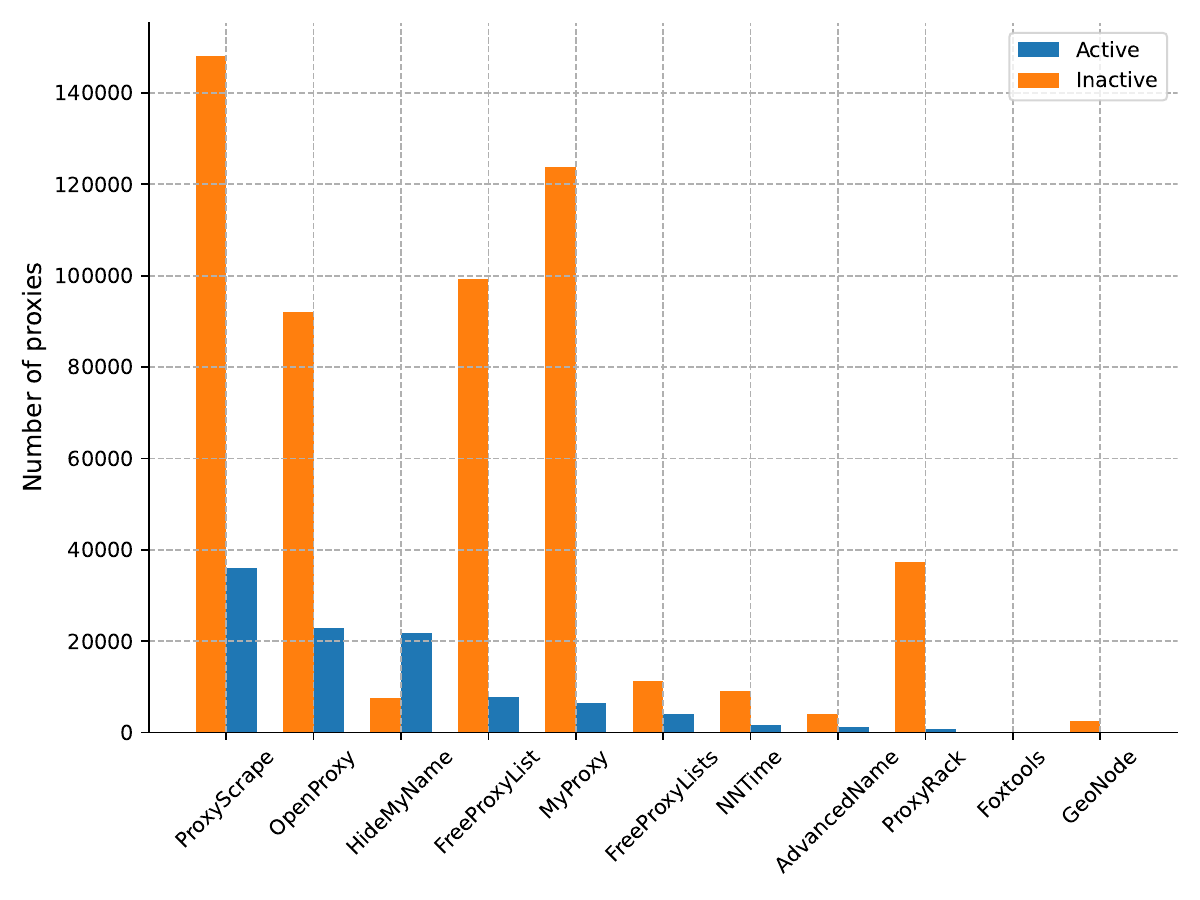}
    \caption{For each provider, the number of active and inactive proxies on the first test after collection.}
    \label{fig:firsttestprovider}
\end{figure}

\parhead{Providers.} Figure~\ref{fig:evproviderunique} shows the evolution of unique proxies for each provider.
In this figure, we only consider proxies that were never seen on previous collections and that are exclusive to one provider when being collected.
It can be inferred that while some proxy providers, such as \textit{MyProxy} or \textit{ProxyScrape} tend to offer unique proxies on a regular basis, others are more likely to either reuse the same proxies they initially provided or use proxies that are shared by multiple aggregators. 
Figure~\ref{fig:firsttestprovider} shows that this variation of quality can also be noticed in the number of working proxies on the first test after they have been collected. For instance, \textit{ProxyScrape} and \textit{OpenProxy} are both two of the biggest providers of unique proxies in our dataset and are also the two highest providers of working proxies on their first test. However, \textit{MyProxy} displays a low rate of working proxies on their first test even though the platform is the most consistent provider of unique proxies for the duration of our study.

\parhead{Active proxies.} We break down SOCKS and HTTP proxies into 5 categories (Table~\ref{table:proxydistro}) according to their level of activity:
\begin{enumerate}
    \item \textbf{Active}, which describes proxies that have been active on over 90\% of the tests.
    \item \textbf{Intermittent} for proxies that have shown activity on 50 to 90\% of the tests.
    \item \textbf{Rarely active} describes proxies that have been active between 10 and 50\% of the tests.
    \item \textbf{Short-lived}, which characterizes proxies that have been active at least once but for no more than 10\% of all tests.
    \item \textbf{Never active}, for proxies that have never passed a test.
\end{enumerate}
\hl{Proxy activity is computed by the ratio of the number of tests a proxy has responded to and its total number of tests. For example, if a proxy is tested 5 times and has been active for 3 of these, it will be classified in the \textit{Intermittent} category.} \naif{Following R1's comment}

\begin{table}[t]
    \centering
    \renewcommand{\arraystretch}{1.3} 
    \caption{Distribution of proxies by protocol and activity}
    \label{table:proxydistro}
    \begin{tabular}{|>{\centering\arraybackslash}p{1.5cm}|>{\centering\arraybackslash}p{1cm}|>{\centering\arraybackslash}p{1cm}|>{\centering\arraybackslash}p{1cm}|}
    \hline
    \textbf{Protocol}       & \textbf{HTTP}   & \textbf{SOCKS}   & \textbf{Total}  \\ \hline
    Active                  & 12,199                                 & 70                                & 12,269                              \\ \hline
    Intermittent            & 6,255                                  & 189                               & 6,444                               \\ \hline
    Rarely active           & 12,724                                 & 6,293                             & 19,017                              \\ \hline
    Short lived             & 78,984                                 & 104,605                           & 183,589                             \\ \hline
    Never active            & 258,466                                & 160,908                           & 419,374                             \\ \hline
    \textbf{Total}          & \textbf{368,628}                       & \textbf{272,065}                  & \textbf{640,693} \\ \hline
    \end{tabular}
\end{table}

Our dataset includes a total of $368,628$ HTTP proxies and $272,065$ SOCKS proxies.
Our results show that HTTP proxies outperform SOCKS proxies in terms of stability as the latter only have 70 proxies that have been active on over 90\% of the tests, as opposed to $12,199$ for HTTP proxies.
In their study, Mani~\etal~\cite{mani2018extensive} find comparable results, showing that there are fewer active SOCKS proxies compared to their \textit{HTTP} counterparts. It should be noted that SOCKS proxies forward arbitrary TCP and UDP traffic allowing TLS connections to be established with the requested websites. However, HTTP proxies can avoid supporting HTTPS connections, as most of them do, forcing users to use plain HTTP and eavesdrop or alter their network traffic.

\begin{figure}[t]
    \centering
    \includegraphics[width=1\columnwidth]{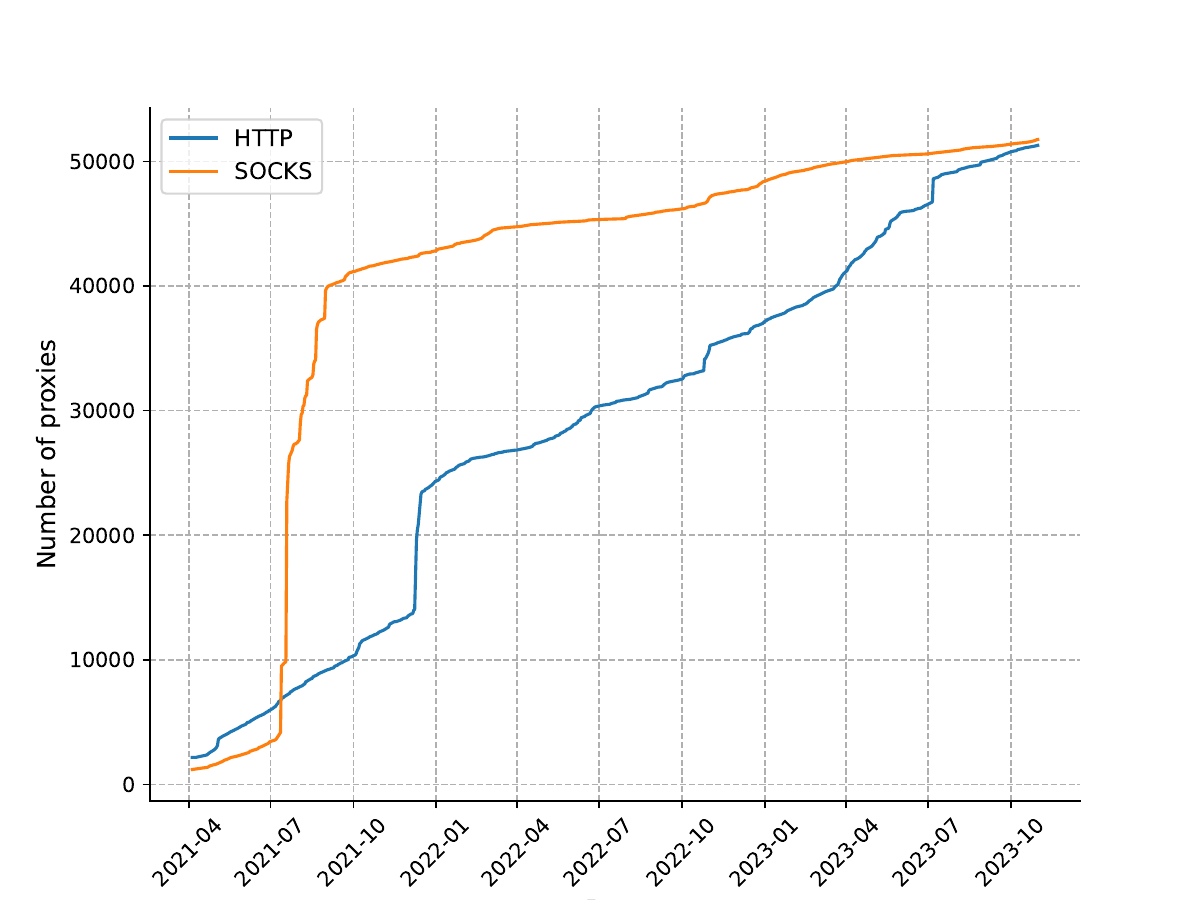}
    \caption{Cumulative sum of HTTP (in blue) and SOCKS (in orange) proxies that are active on their first test.}
    \label{fig:firsttest}
\end{figure}

Figure~\ref{fig:firsttest} shows the cumulative number of proxies that are active on their first test, for each protocol.
Both protocols have a popularity spike at different periods: 
\begin{itemize}
    \item SOCKS proxies are the first to display a significant increase that starts in July 2021. 
    \item HTTP proxies follow the same trend, albeit with a significantly lower increase, in December 2021.
\end{itemize}
We note that the surge in active SOCKS proxies correlates to a corresponding increase in collected SOCKS proxies. We do not have an explanation for this increase.
However, in the case of HTTP proxies, the increase can potentially be associated with the \textit{Log4j} vulnerability~\cite{log4jlunasec} (CVE-2021-44228) that was made public on December 10th, 2021. The \textit{Log4j} vulnerability allowed attackers to execute arbitrary code on servers or computers and was identified in the Java \textit{Log4j} library. It was estimated that hundreds of millions of devices were affected by the vulnerability~\cite{hiesgen2022race}. 
Recently, Sysdig's Threat Research team has identified \textit{proxyjacking} attacks that leverage the \textit{Log4j} vulnerability to install a proxy server and sell the compromised IP address to proxy providers~\cite{morin2023log4j}. We assume that a share of proxies in our list exists due to the exploitation of the \textit{Log4j} vulnerability.

\begin{figure}[]
    \centering
    \includegraphics[width=1.0\columnwidth]{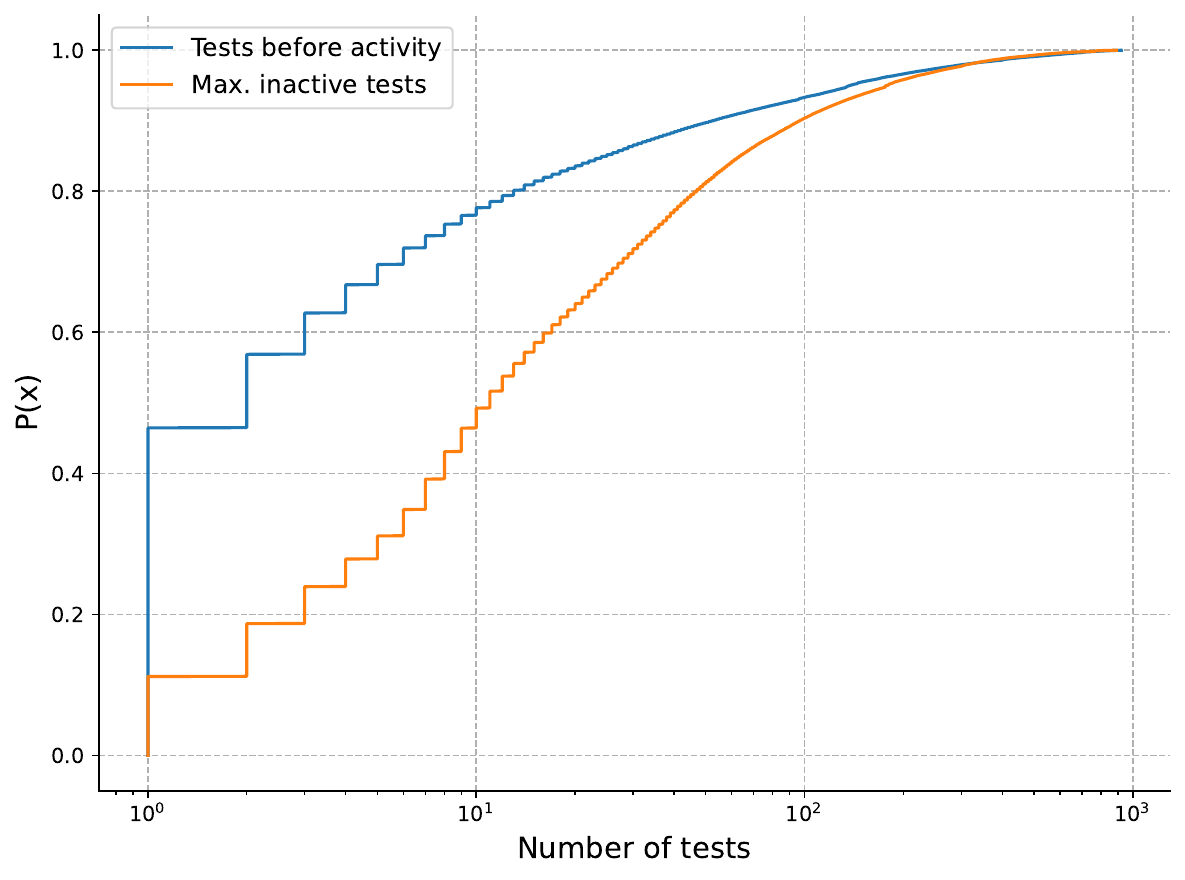}
    \center
    \caption{In blue, out of $221,319$ proxies that were active at least once, the CDF of the number of tests before a proxy became active. After becoming inactive, $156,015$ proxies become active at least once more. In orange, the CDF of the maximum number of successive inactive tests before becoming active again.}
    \label{fig:delaycdf}
\end{figure}

The majority of proxies, regardless of protocol, tend to be active on their first run, as depicted in the \textit{cumulative distribution function (CDF)} of the number of tests a proxy takes to be found active after collection (Figure ~\ref{fig:delaycdf}). It can be noted that while $46.4$\% of proxies that showcased activity through their testings are alive on their first test, at least $22.3$\% of collected proxies took over 10 days before activating. This behavior can be explained by multiple factors: due to the ease of accessing free proxy lists, proxies are often faced with significant loads, rendering them too slow to answer before the timeout (\cref{sec:methodo}) at the moment they are requested.
Another possibility is that proxies might be installed on devices that do not remain constantly active, therefore only responding to probes when the device is running.

\begin{figure}[t]
   \centering
   \includegraphics[width=1.0\columnwidth]{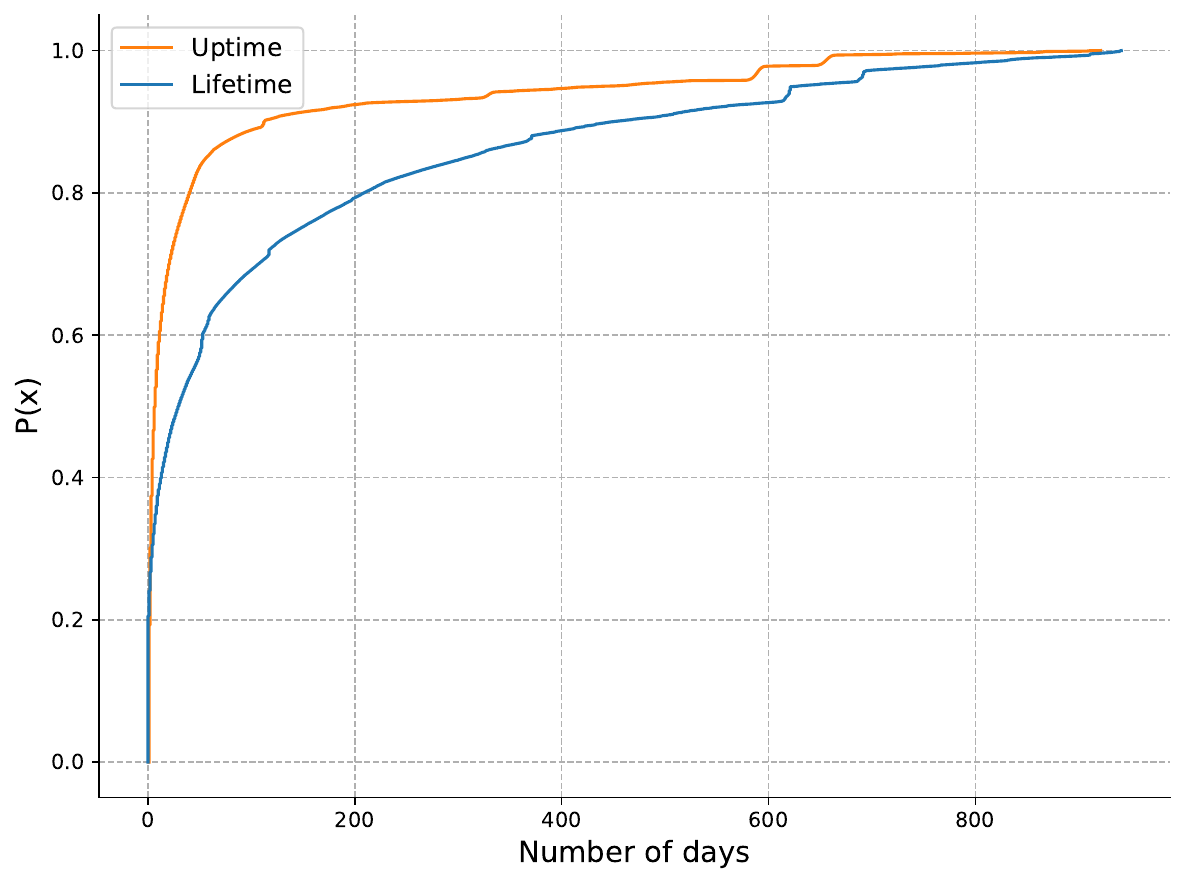}
   \caption{CDF of the lifetime and uptime for each active proxy. The lifetime is the number of days between the first and last time the proxy was active. The uptime is the number of days the proxy was active.}
   \label{fig:uptimecdf}
\end{figure}

Similarly to Perino~\etal~\cite{perino2018proxytorrent}, we define the \textit{uptime} as the number of days a proxy was active within its lifetime, and the \textit{lifetime} as the number of days between the first and last time a proxy has been seen active.
We present in Figure~\ref{fig:uptimecdf} the CDFs of the lifetimes and uptimes of the proxies over 30 months. 
Similarly to Perino~\etal, we observe that proxies tend to have a long lifetime, with \hl{$30.8$\%} of proxies lasting over 100 days and \hl{$9.1$\%} over 500 days.
The proxies' uptime, however, is usually short, with almost $50$\% displaying an uptime of only \hl{6} days or more. \hl{Moreover, only $11$\% of proxies showcase an uptime of over 100 days.}
Despite a few proxies being stable, these results confirm that, in general, stability cannot be expected in free proxies. 

The majority of open proxies in our database tend to be unstable, with cycles of activity and inactivity.
For each proxy that cycled through an inactive period in our database (i.e., $156,015$ proxies), in Figure~\ref{fig:delaycdf} we compute the maximum number of continuous tests a given proxy fails before becoming active again.
The CDF shows that even though proxies might look inactive for a continuous period of time, it cannot be assumed the proxy is dead.
We note that over $50.8$\% of such proxies did not respond to our probes for up to 10 days before resuming activity, and $10.8$\% of proxies were inactive for up to 90 days before resuming.
In more extreme cases, \hl{$14,970$ (9.5\%)} proxies were inactive for over 100 days before resuming: $1,085$ of those had been inactive for over 500 days, $594$ for over 600 days and $110$ for over 800 days.
On average, when a proxy stops working, it takes up to $39$ days before it resumes activity. 
\hl{Finally, out of the proxies that were tested multiple times, we identify $42,515$ proxies that worked only once through their lifetime. This number is significantly lower than the number of proxies that cycle through active and inactive states, further showcasing the low stability and reliability of free proxies.}

\parhead{Geographical distribution}

To gain a further understanding of the free proxy ecosystem, we collect the source country and autonomous system of each of the collected proxies, following the methodology described in Section~\ref{sec:methodo}.
Similarly to previous works~\cite{choi2020understanding,tsirantonakis2018large}, we find that the USA and China represent the top 2 providers of open proxies, with Indonesian and Brazilian proxies following, albeit in smaller proportions. 
The reasons for this skewed distribution of proxies toward the USA and China can be explained by two factors: first, these two countries are among the biggest hosters of datacenters in the world~\cite{daigle2021data}, concentrating 33\% for the USA and 4\%  for China. 
Second, it can be noted that the top five countries that provide the most free proxies all have a highly digitalized population, most of them ranking in the top six digitalized countries in the world~\cite{owidinternet}. 
This high proportion of Internet users exposes more connected devices, therefore leading to an increased number of potentially compromised devices, which in turn can be turned into proxies.

\begin{figure}[]
    \centering
    \includegraphics[width=1.0\columnwidth]{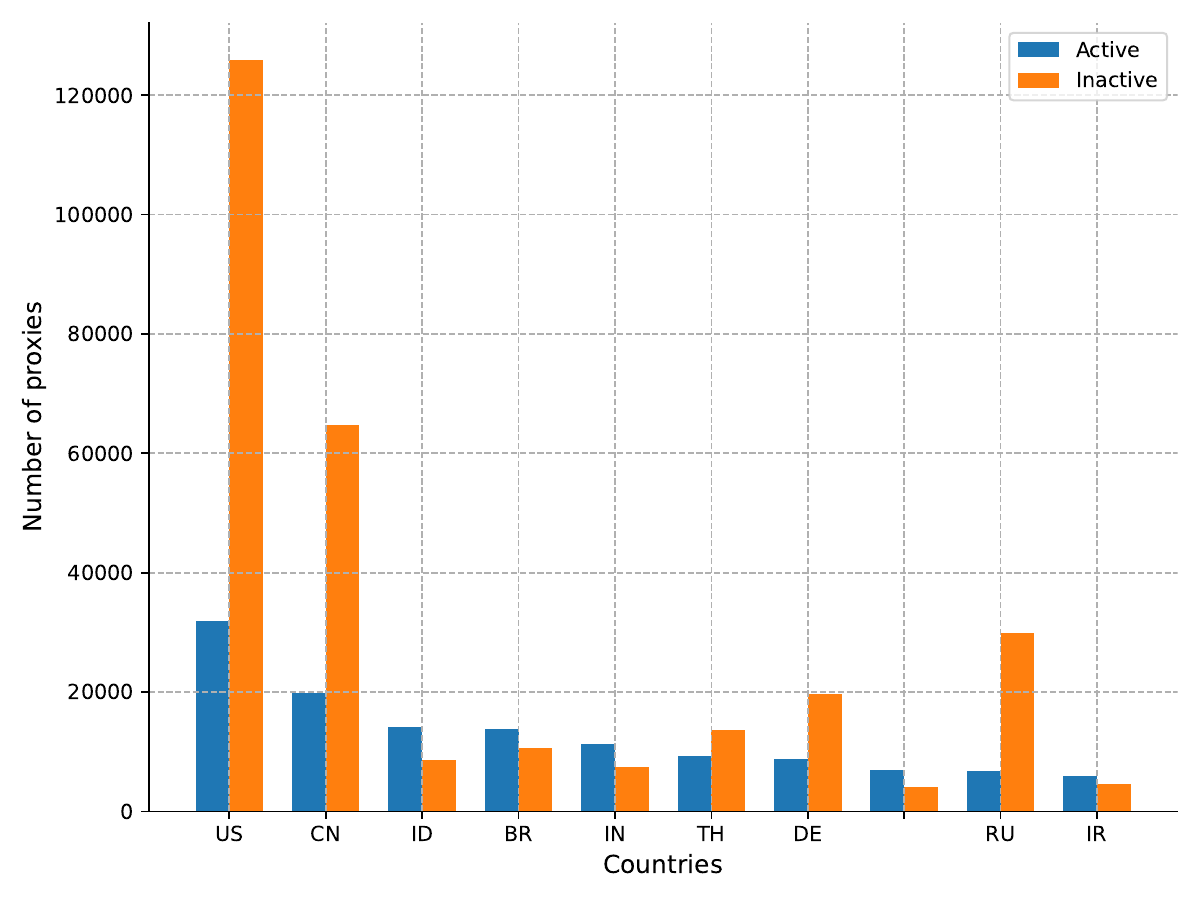}
    \caption{Active and inactive proxies per country (Top 10).}
    \label{fig:countriesactivity}
\end{figure}

Figure~\ref{fig:countriesactivity} presents the proxy activity for the top 10 countries with the most active proxies. We consider a proxy to be active when it has at least one successful test. While the US remains at the first position with $31,906$ active proxies, it is closely followed by China (CN), which provides $19,882$ active proxies. This observation outlines the fact that proxies originating from the US are less guaranteed to be functional than proxies coming from other countries.
 
Table~\ref{table:asndistro} presents the top 10 autonomous systems in our dataset. In total, we count over $15,094$ distinct autonomous systems, with $5,222$ of them being associated with only one proxy in our database. Only $65$ autonomous systems handle more than $1000$ proxies in our database, with \textit{Chinanet} (AS4132) topping the list with over $37,560$ proxies.

\begin{table}[h]
    \caption{Top 10 Autonomous Systems found among the collected proxies}
    \renewcommand{\arraystretch}{1.3}
    \centering
    \begin{tabular}{|l|c|c|}
    \hline
    \multicolumn{2}{|c|}{\textbf{Autonomous system}}   & \multirow{2}{*}{\textbf{\# proxies}} \\ \cline{1-2}
    \textbf{Name} & \textbf{Number}  & \\ \hline
    Chinanet                & AS4132                                   & 37,560                                                           \\ \hline
    Blazing SEO, LLC        & AS397630                                 & 14,244                                                           \\ \hline
    Hosting Solution Ltd.   & AS14576                                  & 11,154                                                        \\ \hline
    Telecom Argentina S.A.  & AS7303                                   & 11,642                                                           \\ \hline
    DigitalOcean            & AS397630                                 & 11,464                                                         \\ \hline
    China Unicom Backbone   & AS4837                                   & 10,905                                                         \\ \hline
    ColorCrossing           & AS36352                                 & 10,597                                                         \\ \hline
    Cloudflare London       & AS209242                                 & 10,226                                                         \\ \hline
    Amazon                  & AS16509                                  & 8,373                                                        \\ \hline
    \end{tabular}
    \label{table:asndistro}
\end{table}

\textit{\textbf{Takeaway:} The free proxies announced on publicly available lists are plentiful but only a small fraction are effectively active at a given time. 
Our data collection over 30 months shows how unstable and unreliable most free proxies are and the majority of them are located in the US and China.}

\subsection{Security}

\hl{The results obtained in Section~\ref{subsec:activity} show that most free proxies are unstable and unreliable. Nonetheless, there are still cases of highly reliable proxies, despite them being free, leading us to question the reasons motivating the creation of free proxies. In this section, we leverage the Shodan database to explore whether the presence of vulnerable software and hardware could contribute to the emergence of a free proxy within a given network.}

\label{sec:shodan_analysis}

\begin{figure}[t]
    \includegraphics[width=1.0\columnwidth]{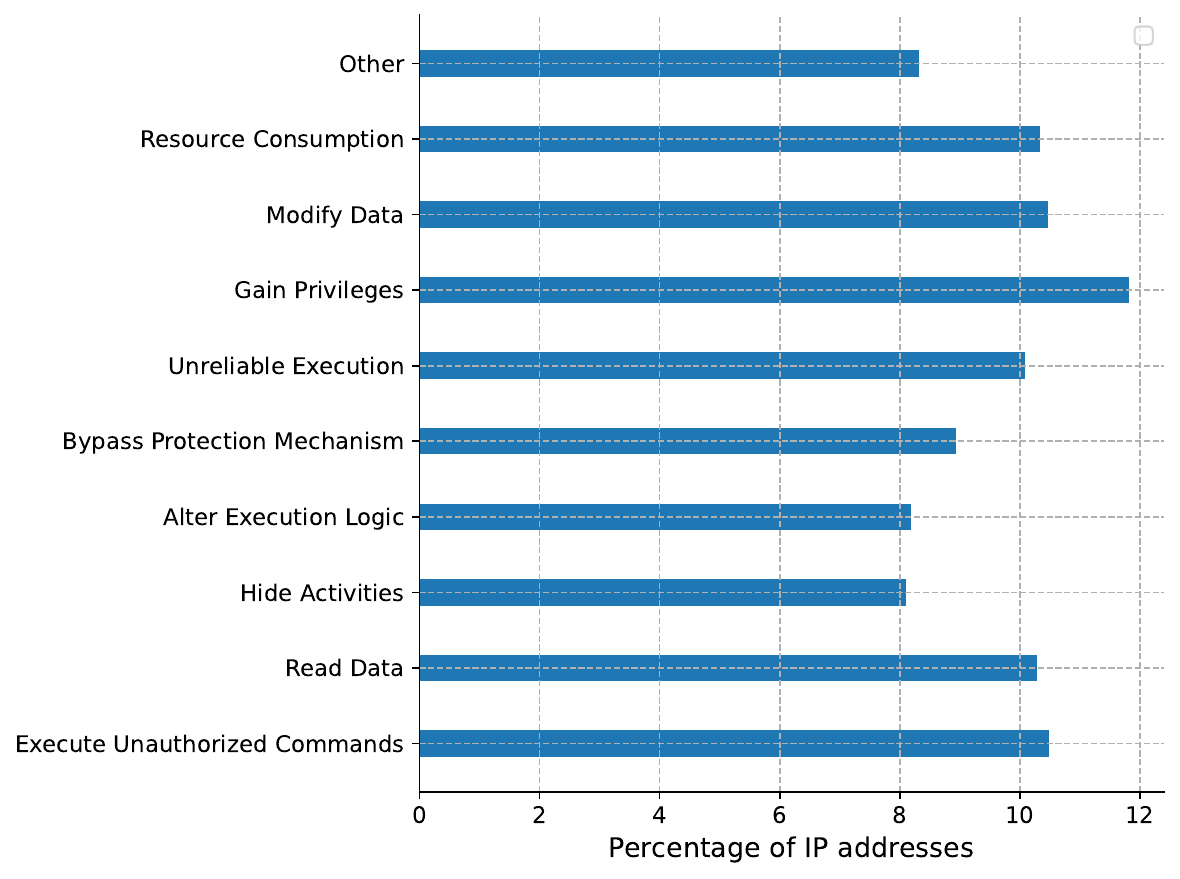}
    \centering
    
    \caption{Percentage of IP addresses per CAPEC impact. One vulnerability can be associated with multiple impacts.}
    \label{fig:capecimpact}
\end{figure}

\parhead{Vulnerabilities.} To present an accurate picture of the free proxy ecosystem, we consulted Shodan for the proxies that were active at least once during our tests. \hl{As mentioned in Section~\ref{subsec:shodan_collection}, the Shodan information for all IP addresses was collected once at the end of the study.}
The Shodan database provides information for $112,623$ of the active proxies' IP addresses \hl{(representing a total of $60$\% of the IP addresses with active proxies)}.
For each service running on a specific port, the Shodan crawler identifies the software powering the service, along with the associated vulnerabilities for the detected version. Overall, the Shodan crawlers identified $4,452$ different CVEs in the detected services and $1,158$ different software. \hl{$13,742$ IP addresses, belonging to $27,371$ proxies that were active at least once have at least one vulnerability.} For each CVE, we use the associated CWEs to obtain the corresponding CAPEC entry. We depict the most common impact of the identified CVEs in Figure~\ref{fig:capecimpact}. It can be observed that CVEs that allow potential attackers to gain privileges on the device are the most common, while the unauthorized execution of arbitrary commands is second. Both these impacts are linked to vulnerabilities that could be potentially used to take control of a device and establish a proxy in the device or on the local network.

We further analyze the CVEs per proxy activity \hl{and for the ports on which the proxy is running}, according to the categories defined in Section~\ref{subsec:activity}. The results are presented in Table~\ref{table:topcveperactivity}. One interesting observation is that the top 5 CVEs found in IP addresses pertaining to short-lived proxies are due to vulnerabilities in the Squid\footnote{\url{http://www.squid-cache.org/}} software, which allows the creation of forwarding proxies.
Most importantly, CVE-2022-41318, which is observed $795$ times, exposes the software to unintended memory reads that can potentially reveal cleartext credentials. 
Potential exploits of this CVE might lead to unauthorized actions if administrative credentials are revealed, which might allow the attacker to set up a hidden proxy. 

\begin{table}[t]
    \centering
    \renewcommand{\arraystretch}{1.5} 
    \caption{Top 5 CVEs per proxy activity category}
    \label{table:topcveperactivity}
    \begin{tabular}{|>{\centering\arraybackslash}p{1.5cm}|>{\arraybackslash}p{2.75cm}|>{\arraybackslash}p{1.65cm}|}
    \hline
    \textbf{Activity} & \textbf{Top CVEs (CVSS score)} & \textbf{Software} \\ \hline
    Active & CVE-2021-3618 (7.4) \newline CVE-2021-23017 (7.7) \newline CVE-2019-20372 (5.3) \newline CVE-2019-9513 (7.5) \newline CVE-2019-9516 (7.5) & ALAPACA \newline Nginx \newline Nginx \newline HTTP/2 \newline HTTP/2 \\ \hline
    Intermittent & CVE-2021-3618 (7.4) \newline CVE-2021-23017 (7.7) \newline CVE-2019-20372 (5.3) \newline CVE-2019-9511 (7.5) \newline CVE-2019-9513 (7.5) & ALAPACA \newline Nginx \newline Nginx \newline HTTP/2 \newline HTTP/2 \\ \hline
    Rarely Active & CVE-2015-1788 (4.3) \newline CVE-2015-1789 (4.3) \newline CVE-2015-1790 (5.0) \newline CVE-2015-1791 (6.8) \newline CVE-2015-1792 (5.0) & OpenSSL \newline OpenSSL \newline OpenSSL \newline OpenSSL \newline OpenSSL \\ \hline
    Short Lived & CVE-2021-46784 (6.5) \newline CVE-2022-41318 (8.6) \newline CVE-2021-28116 (5.3) \newline CVE-2021-28651 (7.5) \newline CVE-2021-28652 (4.9) & Squid \newline Squid \newline Squid \newline Squid \newline Squid \\ \hline
    \end{tabular}
\end{table}

\parhead{Identified Software.} 
Often, the software on the device at an IP address can help identify the device's hardware. Through the information provided by the Shodan database, it is possible to infer that most proxies are hosted on routers. 
In fact, software related to MikroTik routers can be found in 5 of the top 10 most commonly identified software. 
MikroTik-related software can be found on $42,206$ IP addresses ($49,971$ active proxies), and in over $66$\% of these, no third-party software was identified, suggesting that proxies on these IP addresses are likely due to software linked to MikroTik. 
Focusing on the proxies running on a MikroTik device, we find that $602$ distinct CVEs can be identified and directly linked to MikroTik-related software. Most notably, 7 of the top 15 CVEs had a base \textit{Common Vulnerability Scoring System (CVVS)} score of at least 7.5, and 6 of those have a CVSS of 9 and over, positioning them in the \textit{critical} category of the CVSS v3 specification~\cite{cvssv3}. For the \textit{critical} CVEs, we identify the related CWEs and the associated CAPEC entries: the most common \textit{critical} CVE (\textit{CVE-2022-37454}), can lead to the unauthorized execution of commands, with a CVSS entry suggesting a low attack complexity. 
Other specific router models are detected, albeit in a lower proportion: we find mentions of the Ubiquiti AirRouter in $403$ IP addresses and the \textit{ZTE H8102E} router in $541$ of them, with a combined total of $22$ identified CVEs for both of these routers.
Even if the exploitability of the identified vulnerabilities remains open, the ease with which it is possible to identify the software running on a router coupled with the existence of numerous CVEs make them an easy target for attackers.
\naif{continue more ?}\WR{maybe some time of conclusion is missing, like are the proxies because the devices are likely hacked?}\pierre{check to see if the last sentence makes sense}

We also identify a significant number of IP addresses hosting webcams. The most common webcam that was identified is the \textit{Hikvision} camera, which is detected in $1,720$ IP addresses ($1,790$ active proxies) and is the only identified software on $859$ of those. 
In order to identify potential vulnerabilities linked to Hikvision cameras, we use the CVEDetails\footnote{\url{https://cvedetails.com}} website, which allows users to search for vulnerabilities pertaining to specific software. We find one critical vulnerability (CVE-2018-6414) with a base score of 9.8, that allows buffer overflows, which in turn, might potentially lead to arbitrary code execution. The CVSS entry for this vulnerability states that the attack can be performed through the network with low complexity, highlighting the exploitability of this CVE. Furthermore, a white paper published by an anonymous security researcher~\cite{watchfulip2021unauthenticated} identified another critical vulnerability in 2021, targeting the majority of Hikvision cameras at the time. The vulnerability is stated to be exploitable without user interaction, only requires access to the HTTP server ports, and permits an attacker to obtain full root access to the device. Therefore, it is possible for this vulnerability to be exploited to establish a proxy in vulnerable HikVision cameras.\\
The second most common webcam that we identified among the proxies is the \textit{Avtech AVN801} camera, identified in $53$ IP addresses ($60$ active proxies) and exclusive to $16$ of them. According to the CVEDetails website, three critical vulnerabilities are associated with this model of Avtech camera: CVE-2013-4980, CVE-2013-4981, and CVE-2013-4982. The vulnerabilities present a CVSS base score of  $9.0$, $9.0$, and $9.8$ respectively. The first two are due to buffer overflows that potentially lead to arbitrary code execution, while the last vulnerability allows an attacker to gain administrative rights without sufficiently verifying their identity. All cited vulnerabilities might lead to the devices being abused as proxies.

Finally, when focusing on the proxy-powering software, by identifying the software specifically running on the proxy's port, we find $19,620$ Shodan banners. Among those, \textit{CloudFlare} is the most present product for working proxies, likely indicating that the majority of those proxies tend to hide their services behind a CloudFlare reverse proxy service. \hl{In this situation, the Cloudflare reverse proxy's IP address would be shared with the users, and their traffic would then be redirected to the underlying proxy, allowing a proxy to operate without revealing its true identity.} Most of the remaining pieces of software are common for creating proxies: we identify \textit{Nginx}, \textit{Squid Proxy}, \textit{Apache HTTP}, and MikroTik's HTTP proxy service. We also identify proxies running on the same port that is identified as Hikvision's camera service in $26$ IP addresses.

\parhead{IP Address concealing.} For the last 233 runs, we configured our Apache server to persist the IP address for each request, along with the HTTP user-agent, in which we included an identifier for each tested proxy. We compared both the observed IP address and the proxy's advertised IP address and found $17,924$ distinct proxies that conceal their true IP address. It is unclear whether the true IP address is hidden from the proxy user, the website's server, or both. However, when created willingly, it might be assumed that proxy owners attempt to conceal their IP address either from the website's server or the proxy's users, or both, to potentially avoid revealing their true identity.

\begin{figure}[]
    \centering
    \includegraphics[width=1\columnwidth]{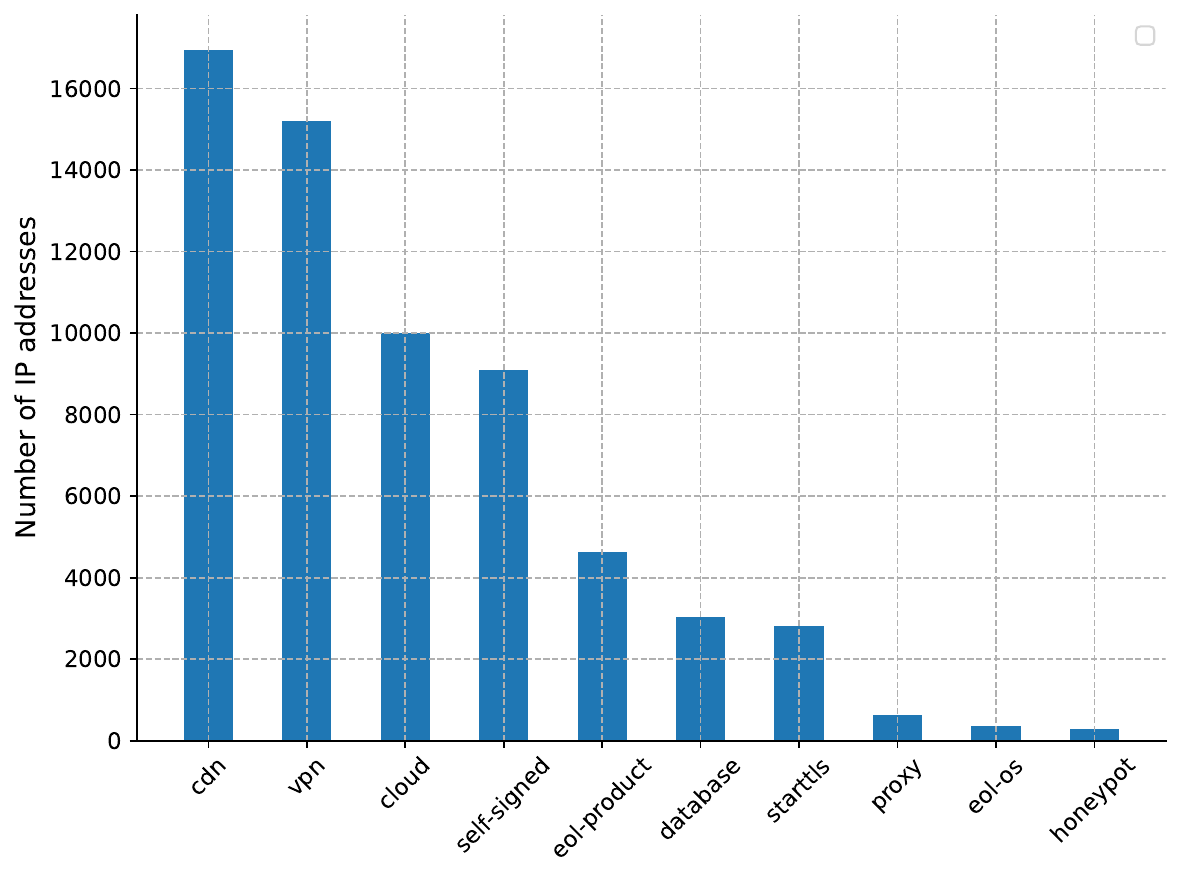}
    \caption{Number of IP addresses associated with Shodan tags (Top 10)}
    \label{fig:tagsactiveip}
\end{figure}

\parhead{Shodan tags.} Shodan's crawlers are configured to detect specific functions of IP addresses, such as whether they host honeypots, or if an IP address belongs to a \textit{Content Delivery Network (CDN)}. Once the Shodan crawlers identify one such function in an IP address, it is tagged with the corresponding attribute.\footnote{\url{https://datapedia.shodan.io/}} 
Figure~\ref{fig:tagsactiveip} depicts the distribution of tags across proxy IP addresses. It can be noted that the majority of IP addresses with a tag are associated with the \textit{CDN} tag, which is related to the presence of \textit{CloudFlare} services in $96.4$\% of cases. Interestingly, we find that \hl{$11,105$} IP addresses under the CDN tag can be classified under the list of \textit{Active} proxies, while only \hl{$5$} of the $15,194$ IP addresses under the \textit{VPN} tag can be considered to be \textit{Active}. Most notably, the vast majority of the proxies for which the corresponding IP address is tagged \textit{VPN} is classified under the \textit{Short lived} category (\hl{$12,716$ IP addresses}). 
One hypothesis that can be emitted is that proxies that are active under IP addresses with the \textit{CDN} tag are intentional proxies that use the CloudFlare service to conceal their true IP address: we find that $10,131$ of the $16,957$ \textit{CDN} IP addresses effectively concealing their real IP address at least once during our tests, suggesting that the proxies are set up intentionally with their identity hidden.\\
It can also be observed that only a small share of IP addresses are attributed to the \textit{proxy} tag, despite being advertised as a proxy by the proxy providers. It is however unclear what reasons lead the Shodan crawlers to appoint the \textit{proxy} tag: most of the identified software on the \textit{proxy}-tagged IP addresses allow the creation and management of proxy services, such as \textit{Nginx} or \textit{Squid}. But proxy-related software is present in multiple other IP addresses with different or no tags at all. \\
Finally, we note that $23$ IP addresses have the \textit{compromised} tag, with the top 3 most identified software being \textit{Nginx}, \textit{Redis} and \textit{OpenSSH}. Out of the top identified CVEs for \textit{compromised} IP addresses, we find two \textit{critical} vulnerabilities, with CVSS scores of 9.8 and both pertaining to \textit{OpenSSH}: CVE-2022-2068 and CVE-2022-1292, which permit an attacker to execute arbitrary code with a low attack complexity.

\parhead{Content modification.}
As mentioned in Section~\ref{sec:methodo}, for each successful test (i.e., active proxy), we verify whether the content returned by the proxy corresponds to the body of our \textit{honeysite}. We find that $16,923$ distinct proxies altered the content at least once throughout our tests.
On average, for each run, $5.7\%$ of active proxies return content that is altered. This ratio is fairly stable in both HTTP and SOCKS proxies, with the first showcasing an average of $6.6$\% of content-altering proxies and the second an average of $6$\%.
\hl{On average, the altered content consists of 36 kilobytes, while the plain text content of our honeysite is 2 bytes. This is due to the majority ($88.9$\%) of content-altering proxies returning full-fledged HTML pages, while the remaining returned plain text. We note that most altered content is due to misconfigured services rather than malicious intent: when extracting all URLs from the persisted HTML pages, we find that multiple of the most common URLs are related to MikroTik, hinting at misconfigured MikroTik routers. Tsirantonakis~\etal~\cite{tsirantonakis2018large} reach the same conclusions and state that only $5.15$\% of proxies that alter the expected content do so with malicious intent. \\
We also identify a significant number of URLs related to cloud providers, such as Tencent or DigitalOcean, which indicates that services were misconfigured on cloud instances. These observations align with the findings obtained through the Shodan database.}

\parhead{Protocols.}
For the last 477 runs of proxy probing, we test the proxies against both the HTTP and HTTPS endpoints of our \textit{honeysite} in order to understand whether they tend to support the secure protocol. Overall, we find that a low portion of free proxies do respond to our probes on the HTTPS endpoint: on average, $371$ proxies were responsive on the secure endpoint, representing an average of $2.5\%$ over the total number of responsive proxies on the HTTP endpoint. Multiple reasons can be advanced regarding this low percentage: first, the required technical understanding is relatively low for setting up a simple proxy without further options as opposed to a proxy that can handle \textit{HTTPS} requests. On top of that, as some free proxies feature malicious behavior~\cite{tsirantonakis2018large}, they might try to avoid handling secure requests as the distributed content is encrypted and tamper-resistant.\\
Some products identified by Shodan also show that the PPTP protocol is present on \hl{$14,249$} IP addresses. PPTP is an outdated protocol that was commonly used in the implementation of \textit{Virtual Private Networks (VPNs)}. However, due to various security issues, it was rendered obsolete by today's standards. PPTP uses the \textit{Microsoft Point-to-Point Encryption (MPPE)}, which has been shown to be easily compromised due to its usage of the RC4 stream cipher~\cite{garman2015attacks}. This can result in an attacker being able to decrypt, analyze, and alter the forwarded traffic carried through PPTP. 
Furthermore, it is possible to use PPTP on both HTTP and SOCKS proxies: one potential motive for using PPTP to create proxies, despite the protocol's vulnerabilities, is to use it and exploit these vulnerabilities to either eavesdrop or alter the traffic. 

\textit{\textbf{Takeaway:} Free proxies present in public lists have a wide variety of security issues.
For the proxies themselves, $27,371$ of them with Shodan entries have at least one vulnerability with some being exploitable to compromise their integrity.
Regarding why these proxies are online, a non-negligible number presents malicious behavior by modifying the content of users' requests due to the use of the insecure HTTP protocol.
Very few proxies support encryption, leaving users vulnerable.
Finally, through the Shodan database, we identify many routers of the MikroTik brand and find that they have been the subject of multiple vulnerabilities. Similarly, we also identify connected devices, such as connected cameras, which present critical vulnerabilities that could potentially be used to create proxies.
It is likely that a large portion of these proxies have been installed through exploits.}






\section{Discussion and limitations}
\label{sec:discussion}

\subsection{Discussion}

\subsubsection{Motives to host free proxies}

Free proxies have existed since the early days of the Internet, offering users a means to circumvent geolocked content and mask their actual IP addresses. Nevertheless, the motivations behind both the providers of lists of free proxies and the operators of free proxies raise questions. Prior to the widespread adoption of encrypted traffic and HTTPS~\cite{felt2017measuring}, free proxies had much more potential to intercept and manipulate the traffic of unsuspecting users. This could potentially lead to advertisement injection, phishing attacks, or even lead unsuspicious users to run malware.
Today, major browsers ensure encrypted traffic and issue warning messages if users attempt to access a website with an invalid SSL certificate or without any certificate. Encryption effectively prevents many of the \textit{Man-In-The-Middle (MITM)} attacks from intercepting and altering network traffic, reducing the malicious capabilities of certain free proxies. 
However, in ProxyTorrent~\cite{perino2018proxytorrent}, Perino~\etal~discovered that despite the widespread adoption of HTTPS at the time of their study, various proxies still exhibited malicious behaviors, injecting advertisements or manipulating SSL certificates. Although major browsers can thwart these malicious actions by enforcing HTTPS and validating certificates, free proxy owners may anticipate users relying on outdated or vulnerable software or devices, where their manipulations can succeed.

In our analysis, and despite the general instability of free proxies, we have also identified highly stable and reliable proxies. It is possible that in some cases such proxies result from network misconfigurations leading to their inadvertent exposure to the Internet. 
Additionally, a significant portion of proxies exist on networks exhibiting a high number of critical vulnerabilities, suggesting that some proxies may be put in place by malicious actors.
Finally, Tosun~\etal~\cite{tosun2021resip} explain that in the case of residential proxies, nodes are established as users download and install software that has partnered with a proxy provider~\cite{hollander2019luminati}. For instance, in the case of Luminati, a provider of paid residential proxy services, the installation of a free VPN (HolaVPN) leads users to become an exit node in the Luminati network~\cite{trend2018shining}. It is possible that similar practices may exist for free proxies. Users might be induced to install malicious browser extensions or desktop applications that covertly function as proxies under the hood.

\subsubsection{Vulnerabilities in connected devices}

Multiple vulnerabilities have made devices susceptible to compromise by attackers. In recent years, the \textit{Log4j} vulnerability (CVE-2021-44228) has been shown to affect millions of devices~\cite{hiesgen2022race} and allow attackers to execute unauthorized code with root privileges. The \textit{Log4j} vulnerability is only one among many potentially critical vulnerabilities affecting a constantly growing number of users. With the rising popularity of connected devices, overseeing and controlling the security implications of the software they run becomes a challenging task. Some devices might be sold with outdated and vulnerable software, placing non-technically savvy users at risk of exploitation. The Shodan database provides various examples of such vulnerable devices: \textit{Industrial Control Systems (ICS)} are readily accessible when filtering on the \textit{ics} tag and openly accessible private connected cameras are also documented.
Through our analysis, we identified over $4,500$ vulnerabilities in the IP addresses used by the proxies we tested.
However, many vulnerabilities still remain unidentified since the Shodan database is incomplete.

\subsection{Limitations}

We acknowledge two main limitations in our study: first, due to scaling factors, we only test each proxy once a day. At the time of testing, the proxy might be under heavy load or temporarily unavailable, preventing it from forwarding our requests and therefore classified as inactive for a given run. Testing the proxies multiple times a day would help mitigate this limitation.

While the Shodan database provides a rich overview of the services running on a given IP address, the Shodan crawlers only scan a specific list of ports, while hosted proxies or vulnerable software might be running on any port. To help address this limitation, future works can consider pulling security-related data from various sources, such as \textit{Censys} or \textit{Hunter.how}, or performing the scans themselves.

\subsection{Ethics considerations}

Our study was evaluated and pursued based on the ethical principles listed in the Menlo Reports~\cite{dittrich2012menlo}. We crawled proxy providers once a day and did not repeat failed requests. All our tests sent through the proxies were performed against domains under our control, through publicly available proxies. We did not overload any tested proxy as we performed two daily requests at maximum. Vulnerability information and scanned services were collected through the publicly available Shodan database. We did not perform any port scanning or any activity that might negatively expose the targeted proxy's IP address.

\section{Conclusion}
\label{sec:conclusion}
In this work, we provide an overview of the free proxy ecosystem and the services that power them. Through the collection of over $640,600$ free proxies over 30 months, and our analysis of the Shodan dataset, we are able to identify $4,452$ vulnerabilities among $1,158$ distinct software present on the proxies' networks. The results show significant security risks in free proxies that can be exploited to turn devices into unwanted proxies.

Through our longitudinal analysis, we highlight multiple aspects of the free proxy ecosystem: we show that free proxies are mainly unstable and unreliable, and pose a threat to regular users due to content manipulation and insecure protocols. We note however that some free proxies are stable and function as expected.
Our findings also reveal several implications. The widespread vulnerabilities and software with proxy capabilities indicate many free proxies likely result from compromised devices rather than intentional deployments. The intentional use of CloudFlare and IP concealment by some proxies suggest efforts to mask the servers' true locations, often indicating a willingness to pursue malicious intent. For users, the major risks around privacy, content manipulation, and lack of secure protocols like HTTPS underline why free proxies should generally be avoided despite promises of anonymity.

Future work can build on this study by further analyzing the nature of content manipulation, correlating scans over time with emerging exploits, and exploring the mechanisms that transform vulnerable hosts into proxies. With the growing adoption of the Internet of Things, it is likely free proxies based on compromised hosts will remain prevalent for the foreseeable future. Ultimately, curbing this ecosystem hinges on improving consumer awareness and demanding better baseline security from device manufacturers.

\section{Acknowledgments}

This work has been financially supported by the \textit{Agence Nationale de la Recherche} through the ANR-19-CE39-00201 FP-Locker\footnote{\url{https://anr.fr/Projet-ANR-19-CE39-0002}} and ANR-21-CE39-0019 FACADE\footnote{\url{https://anr.fr/Project-ANR-21-CE39-0019}} projects, and was made possible by \textit{Software Heritage}, \footnote{\url{https://www.softwareheritage.org/}} the great library of source code.


{\footnotesize 
\bibliographystyle{acm}
\bibliography{free-proxies}}

\begin{thebibliography}{10}

\bibitem{acar2014web}
{\sc Acar, G., Eubank, C., Englehardt, S., Juarez, M., Narayanan, A., and Diaz,
  C.}
\newblock The web never forgets: Persistent tracking mechanisms in the wild.
\newblock In {\em Proceedings of the 2014 ACM SIGSAC conference on computer and
  communications security\/} (2014), pp.~674--689.
\newblock \url{https://doi.org/10.1145/2660267.2660347}.

\bibitem{albataineh2019iot}
{\sc Albataineh, A., and Alsmadi, I.}
\newblock Iot and the risk of internet exposure: Risk assessment using shodan
  queries.
\newblock In {\em 2019 IEEE 20th International Symposium on" A World of
  Wireless, Mobile and Multimedia Networks"(WoWMoM)\/} (2019), IEEE, pp.~1--5.
\newblock \url{https://doi.org/10.1109/WoWMoM.2019.8792986}.

\bibitem{bennett2021empirical}
{\sc Bennett, C., Abdou, A., and van Oorschot, P.~C.}
\newblock Empirical scanning analysis of censys and shodan.
\newblock {\em NDSS Symposium\/} (2021).
\newblock URL:
  \url{https://www.ndsssymposium.org/wp-content/uploads/madweb2021_23009_paper.pdf}
  (Accessed on 02.07.2022).

\bibitem{bodenheim2014evaluation}
{\sc Bodenheim, R., Butts, J., Dunlap, S., and Mullins, B.}
\newblock Evaluation of the ability of the shodan search engine to identify
  internet-facing industrial control devices.
\newblock {\em International Journal of Critical Infrastructure Protection 7},
  2 (2014), 114--123.
\newblock \url{https://doi.org/10.1016/j.ijcip.2014.03.001}.

\bibitem{bugeja2018investigation}
{\sc Bugeja, J., J{\"o}nsson, D., and Jacobsson, A.}
\newblock An investigation of vulnerabilities in smart connected cameras.
\newblock In {\em 2018 IEEE international conference on pervasive computing and
  communications workshops (PerCom workshops)\/} (2018), IEEE, pp.~537--542.
\newblock \url{https://doi.org/10.1109/PERCOMW.2018.8480184}.

\bibitem{chiapponi2022badpass}
{\sc Chiapponi, E., Dacier, M., Thonnard, O., Fangar, M., and Rigal, V.}
\newblock Badpass: Bots taking advantage of proxy as a service.
\newblock In {\em Information Security Practice and Experience\/} (Cham, 2022),
  C.~Su, D.~Gritzalis, and V.~Piuri, Eds., Springer International Publishing,
  pp.~327--344.
\newblock \url{https://doi.org/10.1007/978-3-031-21280-2_18}.

\bibitem{choi2020understanding}
{\sc Choi, J., Abuhamad, M., Abusnaina, A., Anwar, A., Alshamrani, S., Park,
  J., Nyang, D., and Mohaisen, D.}
\newblock Understanding the proxy ecosystem: A comparative analysis of
  residential and open proxies on the internet.
\newblock {\em IEEE Access 8\/} (2020), 111368--111380.
\newblock \url{https://doi.org/10.1109/ACCESS.2020.3000959}.

\bibitem{daigle2021data}
{\sc Daigle, B.}
\newblock Data centers around the world: A quick look.
\newblock {\em United States International Trade Commission: Washington, DC,
  USA\/} (2021).
\newblock
  \url{https://www.usitc.gov/publications/332/executive_briefings/ebot_data_centers_around_the_world.pdf}.

\bibitem{dittrich2012menlo}
{\sc Dittrich, D., Kenneally, E., et~al.}
\newblock The menlo report: Ethical principles guiding information and
  communication technology research.
\newblock Tech. rep., US Department of Homeland Security, 2012.
\newblock
  \url{https://www.dhs.gov/sites/default/files/publications/CSD-MenloPrinciplesCORE-20120803_1.pdf}.

\bibitem{felt2017measuring}
{\sc Felt, A.~P., Barnes, R., King, A., Palmer, C., Bentzel, C., and Tabriz,
  P.}
\newblock Measuring $\{$HTTPS$\}$ adoption on the web.
\newblock In {\em 26th USENIX security symposium (USENIX security 17)\/}
  (2017), pp.~1323--1338.
\newblock \url{https://doi.org/10.5555/3241189.3241292}.

\bibitem{cvssv3}
{\sc FIRST}.
\newblock Common vulnerability scoring system v3.0: Specification document.
\newblock \url{https://www.first.org/cvss/v3.0/specification-document}, 2015.

\bibitem{log4jlunasec}
{\sc Free~Wortley, C.~T., and Allsion, F.}
\newblock Log4shell: Rce 0-day exploit found in log4j, a popular java logging
  package.
\newblock Blog, December 2021.
\newblock \url{https://www.lunasec.io/docs/blog/log4j-zero-day/}.

\bibitem{garman2015attacks}
{\sc Garman, C., Paterson, K.~G., and Van~der Merwe, T.}
\newblock Attacks only get better: Password recovery attacks against
  $\{$RC4$\}$ in $\{$TLS$\}$.
\newblock In {\em 24th USENIX Security Symposium (USENIX Security 15)\/}
  (2015), pp.~113--128.
\newblock
  \url{https://www.usenix.org/system/files/conference/usenixsecurity15/sec15-paper-garman.pdf}.

\bibitem{hiesgen2022race}
{\sc Hiesgen, R., Nawrocki, M., Schmidt, T.~C., and W{\"a}hlisch, M.}
\newblock The race to the vulnerable: Measuring the log4j shell incident.
\newblock {\em arXiv preprint arXiv:2205.02544\/} (2022).
\newblock \url{https://doi.org/10.48550/ARXIV.2205.02544}.

\bibitem{hollander2019luminati}
{\sc Hollander, R.}
\newblock How luminati obtains its residential ips.
\newblock
  \url{https://web.archive.org/web/20200815175351/https://luminati.io/blog/luminati-network-sdk},
  2019.

\bibitem{watchfulip2021unauthenticated}
{\sc IP, W.}
\newblock Unauthenticated remote code execution (rce) vulnerability in
  hikvision ip camera/nvr firmware (cve-2021-36260).
\newblock
  \url{https://watchfulip.github.io/2021/09/18/Hikvision-IP-Camera-Unauthenticated-RCE.html},
  2021.

\bibitem{mani2018extensive}
{\sc Mani, A., Vaidya, T., Dworken, D., and Sherr, M.}
\newblock An extensive evaluation of the internet's open proxies.
\newblock In {\em Proceedings of the 34th Annual Computer Security Applications
  Conference\/} (2018), pp.~252--265.
\newblock \url{https://doi.org/10.1145/3274694.3274711}.

\bibitem{mi2019resident}
{\sc Mi, X., Feng, X., Liao, X., Liu, B., Wang, X., Qian, F., Li, Z., Alrwais,
  S., Sun, L., and Liu, Y.}
\newblock Resident evil: Understanding residential ip proxy as a dark service.
\newblock In {\em 2019 IEEE symposium on security and privacy (SP)\/} (2019),
  IEEE, pp.~1185--1201.
\newblock \url{https://doi.org/10.1109/SP.2019.00011}.

\bibitem{morin2023log4j}
{\sc Moring, C.}
\newblock Proxyjacking has entered the chat.
\newblock \url{https://sysdig.com/blog/proxyjacking-attackers-log4j-exploited},
  2023.

\bibitem{perino2018proxytorrent}
{\sc Perino, D., Varvello, M., and Soriente, C.}
\newblock Proxytorrent: Untangling the free http (s) proxy ecosystem.
\newblock In {\em Proceedings of the 2018 World Wide Web Conference\/} (2018),
  pp.~197--206.
\newblock \url{https://doi.org/10.1145/3178876.3186086}.

\bibitem{roesner2012detecting}
{\sc Roesner, F., Kohno, T., and Wetherall, D.}
\newblock Detecting and defending against $\{$Third-Party$\}$ tracking on the
  web.
\newblock In {\em 9th USENIX Symposium on Networked Systems Design and
  Implementation (NSDI 12)\/} (2012), pp.~155--168.

\bibitem{owidinternet}
{\sc Roser, M., Ritchie, H., and Ortiz-Ospina, E.}
\newblock Internet.
\newblock {\em Our World in Data\/} (2015).
\newblock \url{https://ourworldindata.org/internet}.

\bibitem{scott2015understanding}
{\sc Scott, W., Bhoraskar, R., and Krishnamurthy, A.}
\newblock Understanding open proxies in the wild.
\newblock {\em Chaos Communication Camp\/} (2015).
\newblock \url{https://wills.co.tt/bitbucket/squid.pdf}.

\bibitem{tosun2021resip}
{\sc Tosun, A., De~Donno, M., Dragoni, N., and Fafoutis, X.}
\newblock Resip host detection: identification of malicious residential ip
  proxy flows.
\newblock In {\em 2021 IEEE International Conference on Consumer Electronics
  (ICCE)\/} (2021), IEEE, pp.~1--6.
\newblock \url{https://doi.org/10.1109/ICCE50685.2021.9427688}.

\bibitem{trend2018shining}
{\sc TrendMicro}.
\newblock Shining a light on the risks of holavpn and luminati.
\newblock
  \url{https://web.archive.org/web/20200923073848/https://www.trendmicro.com/vinfo/us/security/news/cybercrime-and-digital-threats/shining-a-light-on-the-risks-of-holavpn-and-luminati},
  2018.

\bibitem{tsirantonakis2018large}
{\sc Tsirantonakis, G., Ilia, P., Ioannidis, S., Athanasopoulos, E., and
  Polychronakis, M.}
\newblock A large-scale analysis of content modification by open http proxies.
\newblock In {\em NDSS\/} (2018).
\newblock \url{https://doi.org/10.14722/ndss.2018.23244}.

\end{thebibliography}

\end{document}